# A Novel Bi-LSTM And Transformer Architecture For Generating Tabla Music

Roopa Mayya[1] , Vivekanand Venkataraman[2], Anwesh P R[2] and Narayana Darapaneni[3]

[1] PES University, Bangalore, Karnataka, 560050, India
[2] Great Learning, Hyderabad, Telangana, 500089, India
[3] Northwestern University, Evanston, IL 60208, United States

## Abstract

INTRODUCTION: Music generation is a complex task that has received significant attention in recent years, and deep learning techniques have shown promising results in this field.

OBJECTIVES: While extensive work has been carried out on generating Piano and other western music, there is limited research on generating classical Indian music due to the scarcity of Indian music in machine encoded formats. In this technical paper, methods for generating classical Indian music, specifically tabla music, is proposed. Initially, this paper explores piano music generation using deep learning architectures. Then the fundamentals are extended to generating tabla music.

METHODS: Tabla music in waveform (.wav) files are pre-processed using the librosa library in Python. A novel Bi- LSTM with Attention approach and a transformer model are trained on the extracted features and labels.

RESULTS: The models are then used to predict the next sequences of tabla music. A loss of 4.042 and MAE of 1.0814 are achieved with the Bi-LSTM model. With the transformer model, a loss of 55.9278 and MAE of 3.5173 are obtained for tabla music generation.

CONCLUSION: The resulting music embodies a harmonious fusion of novelty and familiarity, pushing the limits of music composition to new horizons.

## 1. Introduction

The field of music generation using deep learning started in the mid-2010s with Google's Magenta project [1]. AIVA, an AI-based music composition software, facilitates the easy and fast creation of original music by musicians and non-musicians.

There is extensive literature available on generating classical piano, jazz and other Western instrumental music due to the abundant availability of these genres in the form of Musical Instrument Digital Interface (MIDI) files, piano rolls and Music-XML formats. On the other hand, classical Indian music, both Hindustani and Carnatic, has distinct characteristics and nuances that make it challenging to fully represent and encode in MIDI formats and other machine encoded formats. Classical Indian music employs microtones, which are pitches that lie between the standard Western music notes. MIDI, by default, follows the equal-tempered tuning system used in Western music, which lacks the flexibility to represent the intricate microtonal variations in Indian classical music accurately. Classical Indian music also features a wide range of traditional instruments with unique playing techniques and characteristics, such as sitar, tabla, veena, sarod, and mridangam. Translating the subtleties and nuances of these instruments into the limited parameters of MIDI can be challenging.

Due to these challenges, good and abundant sources of classical Indian music are scarce and there is limited research on generating classical Indian music. In this paper, we present a novel methodology for generating classical Indian music, specifically tabla music, from waveform (.wav) files using a Bidirectional LSTM with attention mechanism model and a transformer model.

## 2. Related Work

The paper by Chou et al. proposed a single LSTM layer architecture that generates music notes. The model uses a cross-entropy loss function and stochastic gradient descent optimizer to train on 28,39,786 examples of (length-100 sequences, next note) pairs from around 1000 midi files [2]. On the other hand, Arrais de Souza et al. employed a tied parallel network model featuring Biaxial RNN LSTM, trained for 48 hours on a personal desktop computer [3]. Although the generated songs had limited dynamics and complexity, the model allowed control over the speed of note playback, influencing the intensity of the music. Mao et al. describes the implementation of DeepJ [4], which is an extension of the Biaxial LSTM model trained on a dataset of MIDI music from 23 composers across three major classical periods, including Bach and Tchaikovsky. Oore et al. used a LSTM-based Recurrent Neural Network to model performance data, which was able to skip forward in time to the next note event, allowing it to skip time steps that contained rests or held existing notes [5].

The DeepHear architecture utilizes a 4-layer stacked autoencoder to generate ragtime jazz melodies and harmonize melodies [6]. Trained on Scott Joplin's ragtime music, the model generates output in a 4-measure format. Harmonization is achieved by finding values for the bottleneck hidden layer that match a given melody. The results exhibit ragtime characteristics but lack true counterpoint. Peracha presents a 5-layer Transformer encoder model for music generation. The model utilizes fixed sinusoidal and learned input embeddings, 8 attention heads, and a dimension of 512. Training includes a batch size of 1 and the 1-cycle policy with SGD for 30 epochs. The model outperforms a GRU model with a validation Negative Log-Loss (NLL) of 0.394 compared to 0.936 [7]. Huang et al. employ a language-modelling approach to train generative models for symbolic music representation. The JSB Chorale dataset is used, with music represented as a matrix and serialized in a raster-scan fashion. The authors achieve a Validation NLL of 1.835 for transformer models, outperforming Performance RNN and LSTM models [8].

## 3. Approach

Initially, we explored piano music generation using different variants of LSTM based architectures and a transformer architecture. The insights gained from piano music generation were then extended to generating classical tabla music.

## 3.1. MIDI datasets

The Classical Piano MIDI dataset comprises more than 300 pieces of classical piano music in MIDI format, representing compositions from 25 famous composers including Bach, Beethoven and Mozart. The dataset is publicly available for download at http://www.piano-midi.de/.
The MAESTRO (MIDI and Audio Edited for Synchronous Tracks and Organization) dataset, created by researchers from Google's Magenta team, consists of over 200 hours of high-quality classical piano performances. It combines MIDI data and high-resolution audio recordings, making it valuable for various tasks such as music transcription, performance analysis, and music generation. The MAESTRO-v3.0.0 dataset files are available at https://magenta.tensorflow.org/datasets/maestro

The Tabla taala dataset consists of 561 short pieces of Tabla loops for 8 different `taals', namely 'addhatrital', 'bhajani', 'dadra', 'deepchandi', 'ektal', 'jhaptal', 'rupak' and 'trital'. All the files are in .wav format. The Tabla taala dataset can be downloaded from this link: https://www.kaggle.com/datasets/pranav6670/tabla-taala-dataset

## 3.2. Pre-processing

The 'music21' library in Python offers a comprehensive set of tools for music analysis, manipulation, and generation. To begin with, MIDI files are imported into a 'Stream' object using the 'converter' module, which supports various MIDI formats like '.mid', '.midi' and '.kar'. The next step involves extracting the notes and chords from the 'Stream' object using its 'flat' property. It is important to note that the extracted list may contain non-note events such as tempo changes and pedal adjustments. To address this, the non-note events are filtered out, and only the notes and chords are stored in an array for further processing. Fig.1 illustrates the pre-processing steps required to extract and format the relevant information from MIDI files for neural network models.

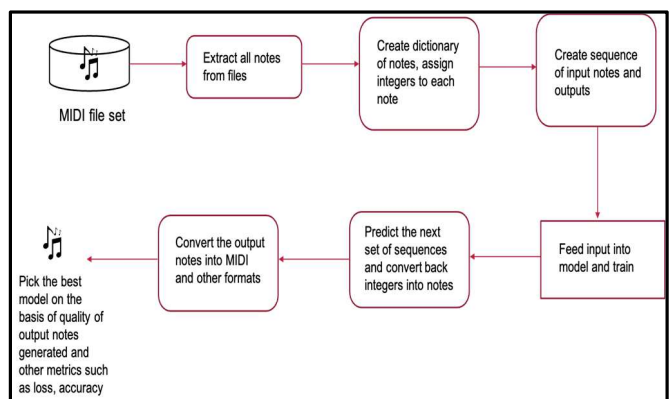

Figure. 1. Process flow diagram of music generation from midi files

To facilitate comprehension by LSTMs, the notes and chords are encoded as unique numerical values using a dictionary. A sliding window approach is employed to create sequences of 100 notes, serving as inputs to train the LSTM. The LSTM predicts the next note and compares it to the actual subsequent note for evaluation.

To pre-process waveform files using the 'librosa' library, the audio data is loaded and converted into a time-domain signal. The signal is then divided into frames, either overlapping or non-overlapping, for further analysis. The frames undergo a Short-Time Fourier Transform (STFT) to obtain the magnitude and phase spectra, often visualized as a mel-spectrogram. Additional representations like mel-frequency cepstral coefficients (MFCCs), spectral contrast, chroma features and tonal centroid can be extracted from the spectrogram. Normalization techniques such as mean normalization or z-score normalization are commonly applied to the extracted features to ensure consistent scales across different audio files.

All the models were trained using a Nvidia A100 GPU using a Google Colab Pro subscription.

# 4. Methodology

Long Short-Term Memory (LSTM) a variant of recurrent neural networks, overcomes the vanishing gradient problem. It is a powerful tool for sequential data analysis, such as in natural language processing (NLP) and music generation. It employs gates and memory cells to selectively retain or discard information, enabling the capture of long-term dependencies while handling noisy input. The gates, using sigmoid and tanh functions, control information flow, while memory cells store and transmit data across time steps.

The Bidirectional LSTM (Bi-LSTM) consists of two LSTM layers that handle input sequences in both forward and backward directions. Each layer's output is combined at each time step before passing through the output layer. This approach captures dependencies in both directions, addressing limitations of a unidirectional LSTM and improving overall performance.

When training on extensive datasets, neural networks can overlook vital information due to the fixed-length context vector structure, resulting in suboptimal performance. To address this, an attention layer is employed to enhance the model's capabilities by focusing on crucial segments of the input sequence when predicting an output. The attention mechanism enables the model to learn associations and extract information from each encoder hidden state, significantly influencing the development of transformers [9]. This mechanism can be seamlessly integrated into neural networks built with different layers. It was initially introduced by Bahdanau et al. in 2014 and has since become an integral part of various architectures.

The alignment scores, $e_{t,i}$, are computed using the encoded hidden states $h_i$ and the previous decoder output, $s_{t-1}$ as denoted in equation (1). These scores represent the alignment between the input sequence elements and the current output at position $t$. A feedforward neural network, represented by a function $a(.)$, can be employed to implement the alignment model.

$$e_{t,i} = a(s_{t-1}, h_i)$$

(1)

The alignment scores obtained earlier are then used to compute the weights through a softmax operation as represented in equation (2):

$$\alpha_{t,i} = \frac{e^{e_{t,i}}}{\sum_{i=1}^{T} e^{e_{t,i}}}$$

(2)

The context vector $c_t$ for the output sequence at position $t$ is calculated using the weighted sum over all the $T$ hidden states as denoted by equation (3):

$$c_t = \sum_{i=1}^{T} \alpha_{t,i} h_i$$

(3)

The transformer architecture, introduced in the 2017 paper "Attention is all you need" by Vaswani et al., incorporates the self-attention mechanism as its core component. This mechanism enables the model to assign importance weights to different words within a sequence [10]. By calculating attention scores between word pairs, the model learns their relevance to each other. The transformer comprises an encoder and a decoder. The encoder processes the input sequence through multiple stacked layers of self-attention and feed-forward neural networks. The decoder generates an output sequence step by step based on the encoded representation.

To compensate for the lack of inherent word order capturing in the transformer model, positional encoding is introduced. It conveys positional information to the model, enabling it to comprehend the sequence's sequential arrangement. The transformer employs multiple attention heads to learn diverse relationships and aspects of the input sequence. The outputs from these heads are combined by concatenation and transformation to yield the final attention representation. Alongside the self-attention mechanism, feed-forward neural networks within each layer further enhance the model's ability to capture complex patterns in the input sequence through non-linear transformations.

To facilitate training of deep networks, the transformer incorporates residual connections, allowing previous layer information to be preserved and aiding learning. Layer normalization is employed after each sub-layer to normalize input and enhance training stability. Masking is applied during training to prevent the model from attending to future words in the decoder. The transformer employs "scaled dot-product attention" for training and generates output tokens sequentially during inference using beam search or greedy decoding strategies.

# 5. Results

First, a set of LSTM variant models were evaluated on a subset of the classical piano midi dataset consisting of 103 songs. The first model comprised of a single LSTM layer, the second model consisted of a LSTM layer and attention layer, the third had a LSTM layer followed by an attention layer and another LSTM layer. The fourth had 3 consecutive layers of an alternating LSTM and attention layer. The fifth model composed of a Bi-LSTM layer followed by an attention layer and another LSTM layer. The sixth model comprised of 3 consecutive layers of an alternating Bi-LSTM and attention layer. Dropout and dense layers were also incorporated in each of these models.

The model comprising of Bi-LSTM, attention and LSTM layers showed the best performance among these 6 models after 40 epochs of training, in terms of loss and accuracy and also the quality of output music generated. Combined graphs of the categorical cross-entropy loss and accuracy for all the 6 LSTM-variant models is shown in fig.2(a) and fig.2(b).

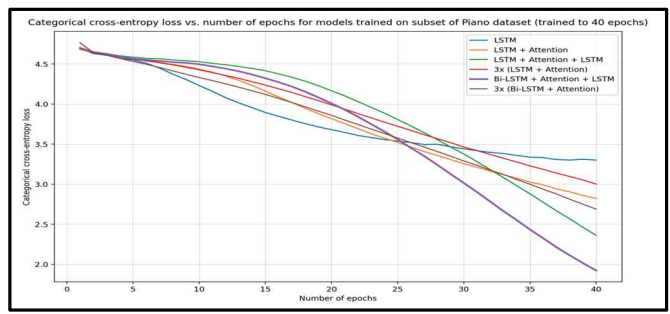

(a)

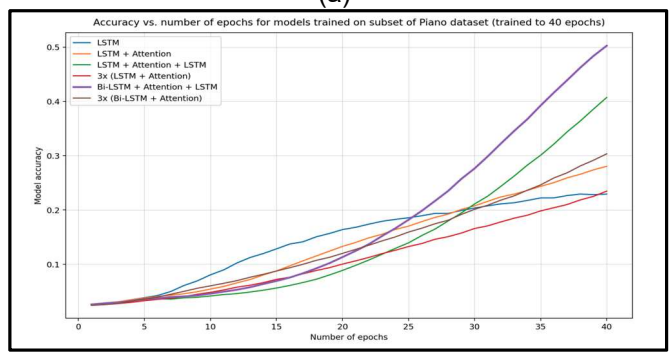

(b)

Fig. 2. Combined graphs of (a) categorical cross-entropy loss and (b) accuracy for models trained on subset of the Classical Piano MIDI dataset

The Bi-LSTM + Attention + LSTM model was then trained for 200 epochs on the full classical piano MIDI dataset of 323 songs. The categorical cross-entropy loss and accuracy plots for this training is shown in fig. 3(a) and fig. 3(b).

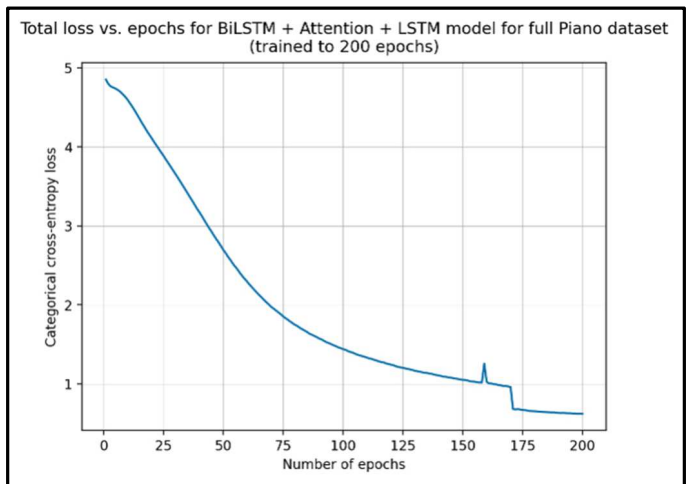

(a)

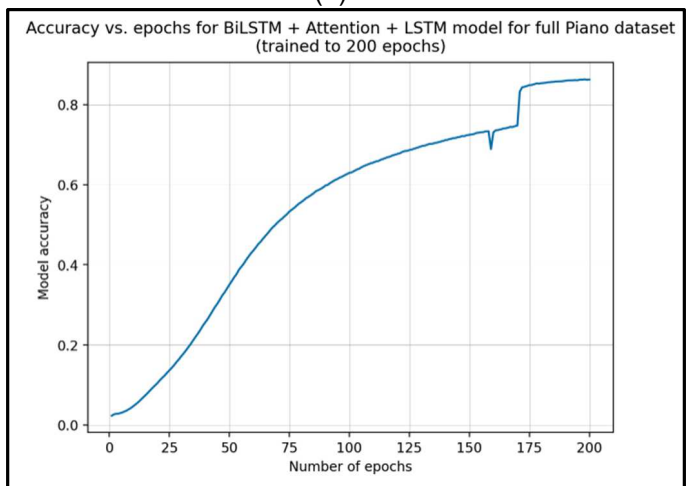

(b)

Fig. 3. Plots of (a) categorical cross-entropy loss and (b) accuracy for the Bi-LSTM + Attention + LSTM model trained on the full classical piano MIDI dataset

Next, a transformer model was evaluated on the full classical piano MIDI and MAESTRO datasets having 1599 songs combined. This model architecture consisted of a token and position embedding layer, followed by multiple transformer-decoder layers with attention heads, and finally, an output layer with a dense layer having units equal to the vocabulary size. The model was trained for 600 epochs and a 'top-p sampler approach' with a 'p' of 0.7 was used for model prediction. The music output of the transformer model was more melodious and had more variations in music notes compared to any of the LSTM variant models.

Table 1 lists the hyper-parameters used for the transformer model for generating piano music.

TABLE 1
HYPER-PARAMETERS USED IN THE TRANSFORMER MODEL
FOR PIANO MUSIC

| Hyper-parameter | Value |
| --- | --- |
| sequence length | 128 |
| embedding dimension | 256 |
| hidden dimension | 256 |
| number of layers | 4 |
| number of attention heads | 8 |
| batch size | 64 |

Graphs of sparse categorical cross-entropy loss and accuracy achieved for the transformer model on piano music are illustrate in fig.4(a) and fig. 4(b).

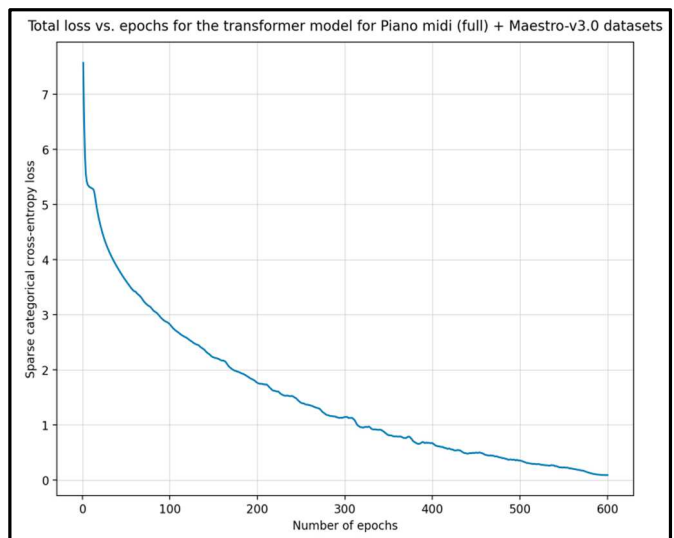

(a)

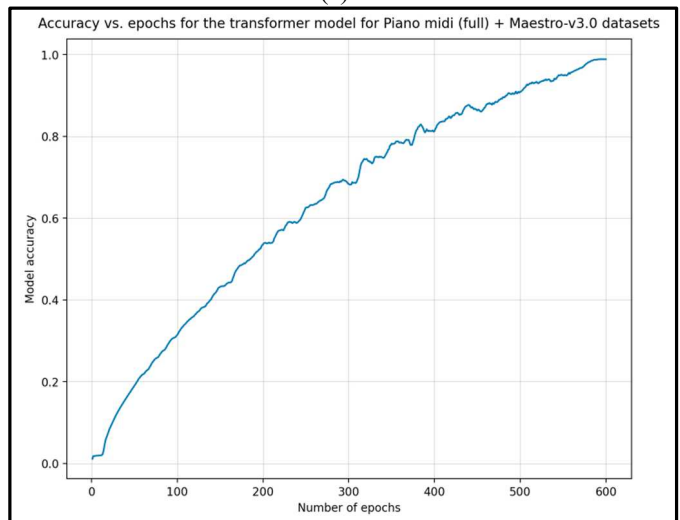

(b)

Fig. 4. Plot of (a) sparse categorical cross-entropy loss and (b) accuracy for the transformer model trained on piano music

The next focus was on extending these concepts to generate tabla music. Few modifications were made to the Bi-LSTM + Attention + LSTM model – instead of one Bi-LSTM layer, two were used. This was followed by an attention layer and 2 LSTM layers. A final dense layer of 128 units was used in the output layer. A 'mean-squared error' loss function and 'Adam' optimizer were used. The model was trained for 300 epochs. The tabla music generated by this Bi-LSTM model was found to be quite comparable and similar to that produced by a human artist.

Fig. 5(a) and 5(b) illustrates the mean-squared error loss and mean-absolute error graphs for the tabla music generation using the Bi-LSTM model.

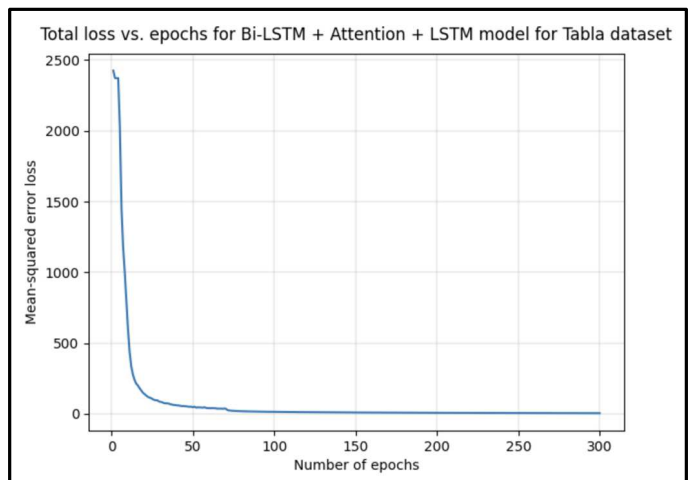

(a)

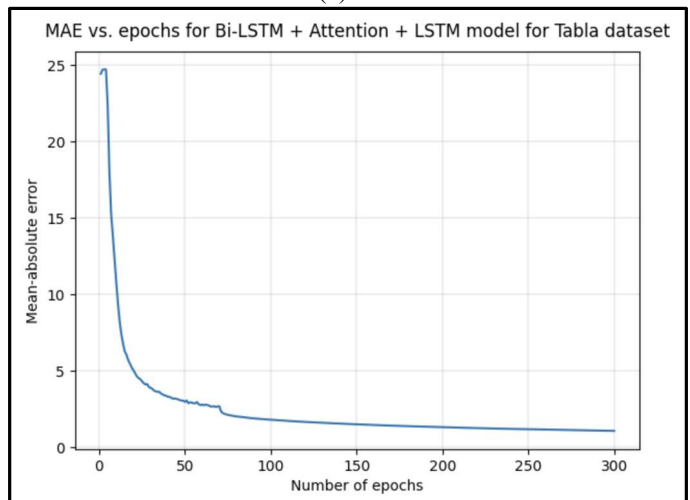

(b)

Fig. 5. Plots of (a)mean-squared error and (b) mean-absolute error for the Bi-LSTM + Attention + LSTM model for generating tabla music

Amplitude vs. time plot and mel-spectrogram of the tabla music generated with the Bi-LSTM model are shown in fig. 6.

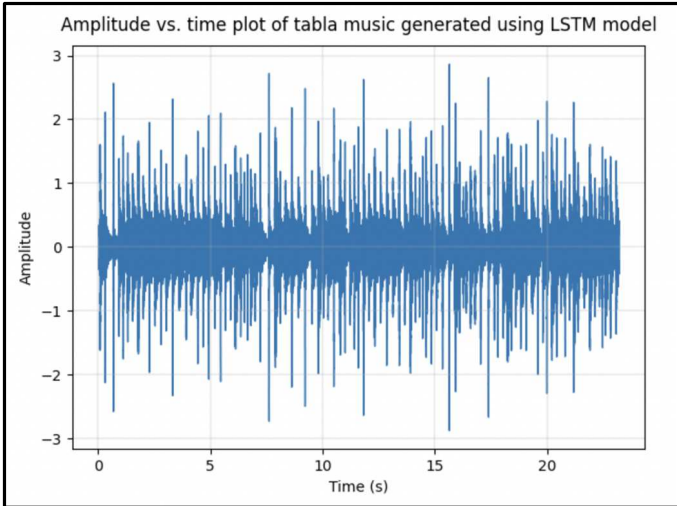

(a)

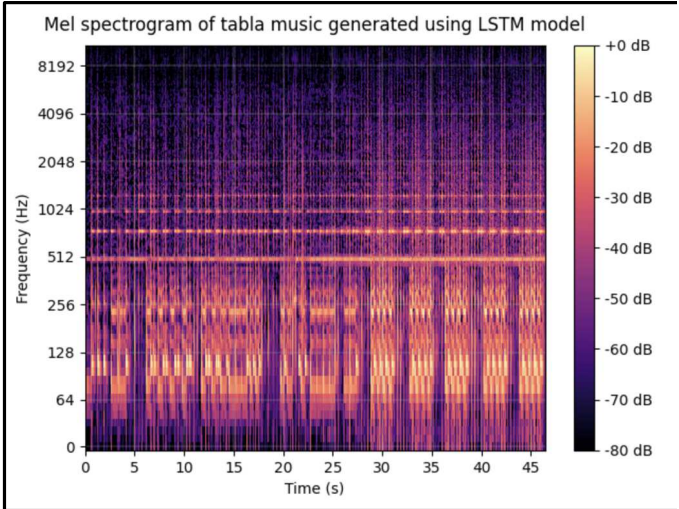

(b)

Fig. 6. Plots of (a) amplitude vs. time and (b) mel-spectrogram of the tabla music generated with the Bi-LSTM model

For the final part of this study, a transformer model was trained on the Tabla Taala dataset. The same pre-processing techniques of the `.wav` files were followed as for the LSTM model. The transformer architecture comprised of a multi-head attention layer followed by dropout, dense and layer normalization layers. These layers were followed by another set of dense, dropout and layer normalization layers. The final output layer was a dense layer of 128 units.

The hyper-parameters used for the transformer model for generating tabla music are listed in table 2.

TABLE 2
HYPER-PARAMETERS USED IN THE TRANSFORMER MODEL FOR TABLA MUSIC

| Hyper-parameter | Value |
| --- | --- |
| sequence length | 60 |
| hidden dimension | 128 |
| number of layers | 6 |
| number of attention heads | 8 |
| batch size | 64 |
| dropout | 0.1 |

The mean-squared error loss and mean-absolute error graphs for the tabla music generation using the transformer model are shown in fig. 7(a) and 7(b).

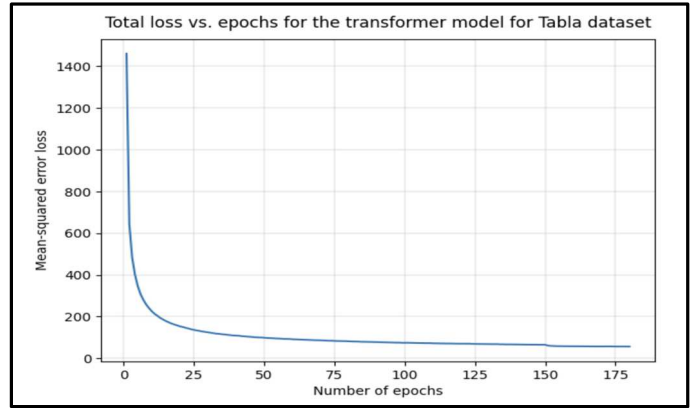

(a)

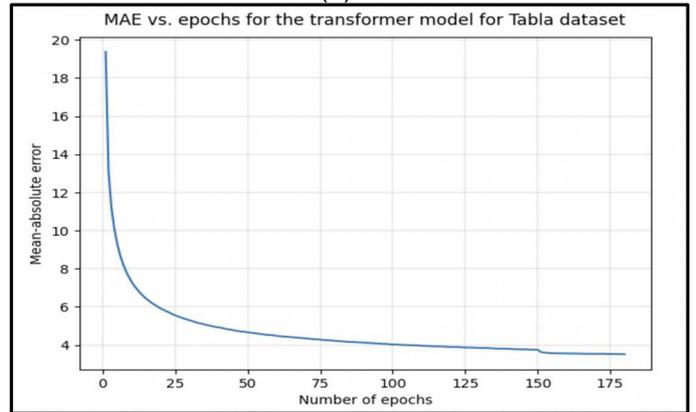

(b)

Fig. 7. Plots of (a)mean-squared error and (b) mean-absolute error for the transformer model for generating tabla music

Fig. 8(a) and 8(b) show the plots for amplitude vs. time plot and mel-spectrogram of the tabla music generated using the transformer architecture.

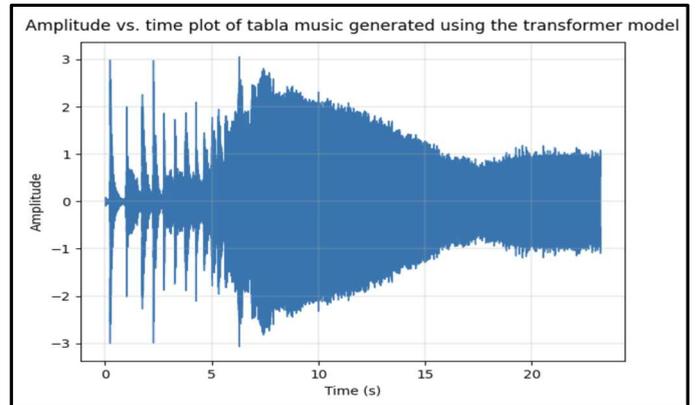

(a)

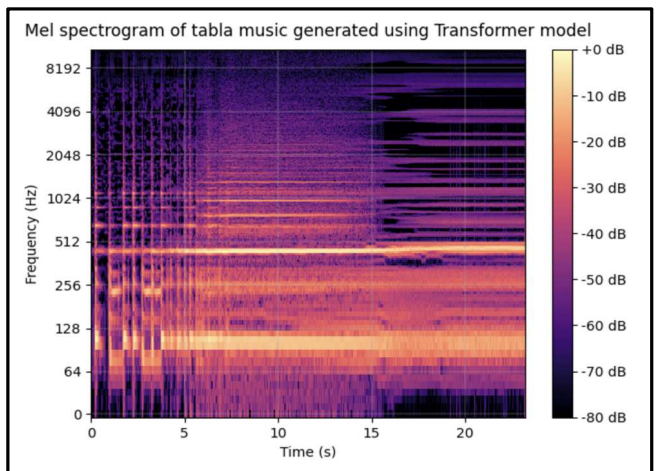

(b)

Fig. 8. Plots of (a) amplitude vs. time and (b) mel-spectrogram of the tabla music generated with the transformer model

Clearly, the transformer model did not perform as well as the Bi-LSTM model for tabla music as evident in the waveplots and mel-spectrograms. Nonetheless, the transformer model was also successful in producing rhythmic tabla sequences for the first few seconds. The model architecture can be refined and trained further to improve the quality of the output tabla sequences.

The observations from all the models discussed in this study are summarized in table 3. A google drive link to all the generated music outputs in this study is provided in the last reference.

TABLE 3

PERFORMANCE EVALUATION OF MODELS FOR PIANO AND TABLA MUSIC GENERATION

| Model | Dataset used | Number of epochs trained | Final loss | Final accuracy / MAE |
|---|---|---|---|---|
| Single-layer LSTM | Classical piano midi (subset) | 40 | 1.9621 | 0.5148 |
| LSTM + Attention | Classical piano midi (subset) | 40 | 1.7975 | 0.4947 |
| LSTM + Attention + LSTM | Classical piano midi (subset) | 40 | 1.9526 | 0.4953 |
| 3x (LSTM + Attention) | Classical piano midi (subset) | 40 | 2.3159 | 0.3768 |
| Bi-LSTM + Attention + LSTM | Classical piano midi (subset) | 40 | 1.2032 | 0.6665 |
| 3x (Bi-LSTM + Attention) | Classical piano midi (subset) | 40 | 2.1492 | 0.4148 |
| Bi-LSTM + Attention + LSTM | Classical piano midi (full) | 200 | 0.6226 | 0.8624 |
| Transformer - Piano | Classical piano midi (full) + MAESTRO | 600 | 0.0920 | 0.9886 |
| Modified Bi-LSTM + Attention + LSTM | Tabla taala | 300 | 4.0427 | 1.0814 |
| Transformer - Tabla | Tabla taala | 180 | 55.9278 | 3.5173 |

## 6. Discussion And Future Work

By extensively training advanced LSTM and transformer models, this study has achieved a novel approach to tabla music generation. The resulting music embodies a harmonious fusion of novelty and familiarity, pushing the limits of music composition to new horizons.

To further develop this study, enhancing the current models by training on a larger tabla dataset is a priority. Additionally, exploring music generation for other classical Indian instruments would be valuable. Another intriguing avenue is generating multi-instrumental music, fusing elements of Indo-Western styles. Lastly, an ambitious endeavour would involve creating a system that mimics human vocal music generation, incorporating lyrics, melody, harmony, and emotional expression. This would require extensive data, computational resources, and training efforts.